# An early explanation of the periodic table: Lars Vegard and X-ray spectroscopy

Helge Kragh[*]

## 1. Introduction

Although found worthy of inclusion in the authoritative *Dictionary of Scientific Biography* [1], the Norwegian physicist and physical chemist Lars Vegard (1880-1963) is not well known outside Scandinavia. His name in the history of science is primarily associated with his pioneering work in auroral research, an interdisciplinary area of science to which he made fundamental contributions. In the period from about 1910 to 1950 he was recognized as the world's foremost authority in auroral spectroscopy [2]. However, in addition to his investigations of the northern light he made significant work also on the borderline between chemistry and physics, in particular as related to X-ray spectroscopy, crystallography, and solid state chemistry. Vegard spent most of his research career at the University of Oslo, from 1913 to 1918 as a docent and from 1918 to 1952 as full professor of physics (Figure 1). In the period 1932-1940 he served as Vice President of the International Union of Physics.

In this paper I focus on Vegard's early attempt, made in a series of papers between 1916 and 1920, to understand the structure of the chemical elements in terms of the electron configurations of atoms. As part of this ambitious research program, in 1918 he suggested configurations of all the elements and on this basis an explanation of the entire periodic system. In fact, his periodic system of that year is probably the first system of its kind, later to be improved by Niels Bohr, Edmund Stoner, and, finally, Wolfgang Pauli. Although a significant contribution to the understanding of the periodic system, one looks in vain for Vegard's name in the standard books on the history of the system, such as Jan Van Spronsen's classical work of 1969 [3] and Erich Scerri's more recent book of 2007 [4]. As noted by Mansel Davies, the importance of Vegard's work in atomic theory "seems to have been very widely overlooked" [5].

## 2. Between Physics and Chemistry

Shortly after having graduated from the University of Oslo (then: Kristiania) in 1905, Vegard became assistant of the physics professor, Kristian Birkeland, who was internationally renowned for his theoretical and experimental work on the aurora borealis. Having received a travel stipend from the Norwegian government, at the end of 1907 he went to Cambridge to study under J. J. Thomson, the famous discoverer of the electron and director of the Cavendish Laboratory. While at Cambridge he published his first scientific work, a series of careful investigations of osmotic properties which attracted the attention of Joseph Larmor, among others [6]. Following postgraduate studies in Cambridge and at the University of Leeds, Vegard went to Würzburg to work in the laboratory of Wilhelm Wien, where he primarily did work on discharges in gases and the positively charged so-called canal rays (atomic or molecular ions), which at the time was a hot topic in physics. It was on this subject he wrote his doctoral dissertation, which resulted in a large paper in the *Annalen der Physik* [7].

During his stay in Würzburg he attended in June 1912 a colloquium in which Max Laue – who was not yet Max von Laue – presented the sensational discovery of diffraction of X-ray in crystals. "Certain new, curious properties of X-rays have been discovered by Dr Laue in Munich," Vegard wrote to W. Henry Bragg in Leeds. "Whatever the explanation may be, it seems to be an effect of most fundamental nature" [8]. His letter of 26 June 1912 triggered the important work in X-ray crystallography of Bragg senior and his son W. Lawrence Bragg that three years later would be rewarded with a Nobel Prize.

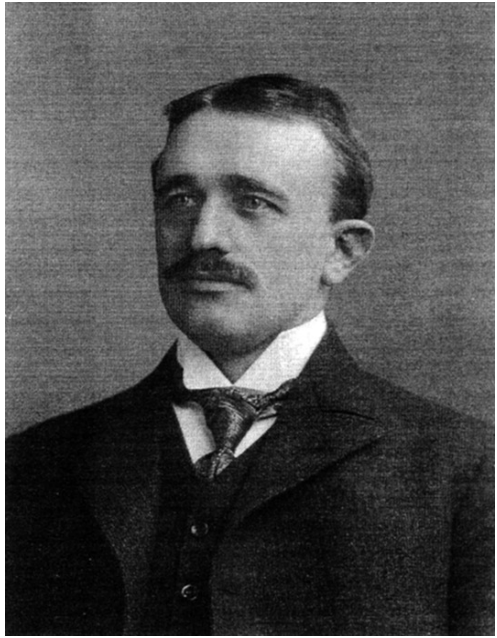

Figure 1. Norwegian physicist Lars Vegard. Figure from Bjørn Pedersen, "Physical science in Oslo," *Physics in Perspective* 13 (2011), 215-238.

After having returned to Oslo, Vegard eagerly took up the new science of X-ray crystallography, while at the same time doing work in auroral research and atomic theory. Among his early works in X-ray crystallography were determinations of the structure of silver, ammonium iodide, rutile ($TiO_2$), and alums which he published in a series of works in the *Philosophical Magazine* [9]. In a paper of 1921 he formulated what is still known as "Vegard's law," an empirical rule according to which the crystal lattice constant of an alloy varies approximately linearly with the concentrations of the constituent elements [10].

This law, and especially the deviations from it, continues to attract attention in mineralogy and materials science.

Throughout his scientific career Vegard continued doing research on structural chemistry, in several cases in relation to his work on the chemical composition of the aurora borealis. For example, in 1924 he suggested – wrongly, it turned out – that the green auroral line of wavelength 5577 Å was caused by minute nitrogen crystals being excited by solar electrons, which led him to investigate the crystal structure of solid nitrogen and other gases in the solid state [11]. The correct explanation of the puzzling green line was established the following year, when John McLennan and his research student Gordon Shrum at the University of Toronto identified it as due to a new transition in atomic oxygen [2].

Vegard's work in crystallography and structural chemistry relied to a large extent on analysis of X-ray spectrograms, and this was not the only use he made of the X-rays. When it came to atomic rather than crystalline structure, it was the characteristic X-ray lines and not the continuous spectrum of the rays that he used as a tool.

### 3. The Bohr Atom and the Periodic System

Although Niels Bohr was not the first to suggest an explanation of the periodic system in terms of arrangements of electrons [12], it was only with his atomic model of 1913 that such suggestions became convincing arguments for the real structure of the system. As Bohr wrote in a letter of 7 February 1913 to his friend, the Hungarian chemist George Hevesy, the still unpublished theory would explain "the way in which the atom-volumes vary with the valence of the element considered" and include "a very suggestive indication of an understanding of the periodic system of the elements" [13]. Bohr's incomplete and tentative explanation, proposed in the second part of his sequel of papers on atomic theory in *Philosophical Magazine*, built on the recently introduced atomic number $Z$ as the ordinal number of the periodic system. Relying on a somewhat arbitrary mixture of physical calculations and empirical data on the physical and chemical properties of the elements, he arrived at electron configurations of the first 24 elements, that is, the number of electrons in the various rings rotating around the central nucleus. For example, he ascribed the structure (8, 2, 1) to sodium, meaning 8 electrons in the innermost ring, 2 in the next ring, and 1 valence electron in the outermost ring.

| | | |
|---|---|---|
| H   1 (1) | F   9 (4, 4, 1) | Cl  17 (8, 4, 4, 1) |
| He  2 (2) | Ne  10 (8, 2) | Ar  18 (8, 8, 2) |
| Li  3 (2, 1) | Na  11 (8, 2, 1) | K   19 (8, 8, 2, 1) |
| Be  4 (2, 2) | Mg  12 (8, 2, 2) | Ca  20 (8, 8, 2, 2) |
| B   5 (2, 3) | Al  13 (8, 2, 3) | Sc  21 (8, 8, 2, 3) |
| C   6 (2, 4) | Si  14 (8, 2, 4) | Ti  22 (8, 8, 2, 4) |
| N   7 (4, 3) | P   15 (8, 4, 3) | V   23 (8, 8, 4, 3) |
| O   8 (4, 2, 2) | S   16 (8, 4, 2, 2) | Cr  24 (8, 8, 4, 2, 2) |

Table 1. Bohr's 1913 proposal of electron rings in chemical elements. In his table in *Philosophical Magazine* he did not assign chemical symbols to the structures.

According to Bohr, the chemical similarity between elements in the same group was a result of the atoms having the same number of electrons in the outermost ring (and not, as J. J. Thomson had earlier suggested, in the inner rings). Thus, he assigned the structure (8, 8, 2, 1) to potassium. Two features with regard to this first quantum-based attempt to reconstruct the periodic system should be emphasized. First, it was provisional and put forward with many reservations. Second, purely physical considerations resulted in some cases in structures that contradicted sound chemical knowledge. In these cases, he opportunistically chose to give higher priority to chemical considerations than mechanical stability calculations. While Bohr had found that the inner ring, to be mechanically stable, could accommodate no more than 7 electrons, in the end he chose the number 8. The reason was obviously the known periodicity of the elements, with the first periods including 8 elements. As to the number of electrons in the outer ring he did not even pretend to base it on calculations: "The number of electrons in this ring is arbitrarily put equal to the normal valency of the corresponding element" [14]. This accounts for the change in the building-up scheme at nitrogen, which he assigned the configuration (4, 3) rather than (2, 5). He gave no reason for this change except that three outer electrons are necessary to account for nitrogen's tervalency.

Although Bohr did not assign electron arrangements to atoms heavier than chromium, based on the periodic system he suggested that "elements of higher atomic weight contain a recurrent configuration of 18 electrons in the innermost rings." Moreover, he argued that in some cases, such as the rare earth metals, the building up of electrons took place in an inner rather than the outer ring. In this way it would be possible to account for the striking chemical similarity of this group of elements. Finally he indicated an explanation of the "observed increase of the electropositive character for an increase of atomic weight of the elements in every single group of the periodic system," say from beryllium to radium. According to Bohr, this was a result of the increasingly weaker binding of the outer electrons as the number of rings increased.

The first one to exploit systematically the chemical potentials of Bohr's atomic theory was Walther Kossel, a young Munich physicist who in 1914 explained the emission of the characteristic X-rays on the basis of the Bohr atom [15], as indicated in Figure 2. According to Kossel, the high-energy K$_\alpha$ line arose from a transition from the L ring ($n$ = 2) to the innermost K ring ($n$ = 1), and K$_\beta$ from a transition from the M ring ($n$ = 3) to the K ring. Similarly, the weaker L radiation was due to transitions $n > 2$ filling a vacancy in the L ring. In an unusually long article in the *Annalen der Physik* from 1916, Kossel extended Bohr's ring structure model to higher elements by connecting the appearance of X-ray series with the emergence of new periods of elements. In building up electron structures, he assumed that "The next electron, which appears in the heavier element, should always be added at the periphery [and] in such a manner that the observed periodicity results" [16]. Kossel elaborated:

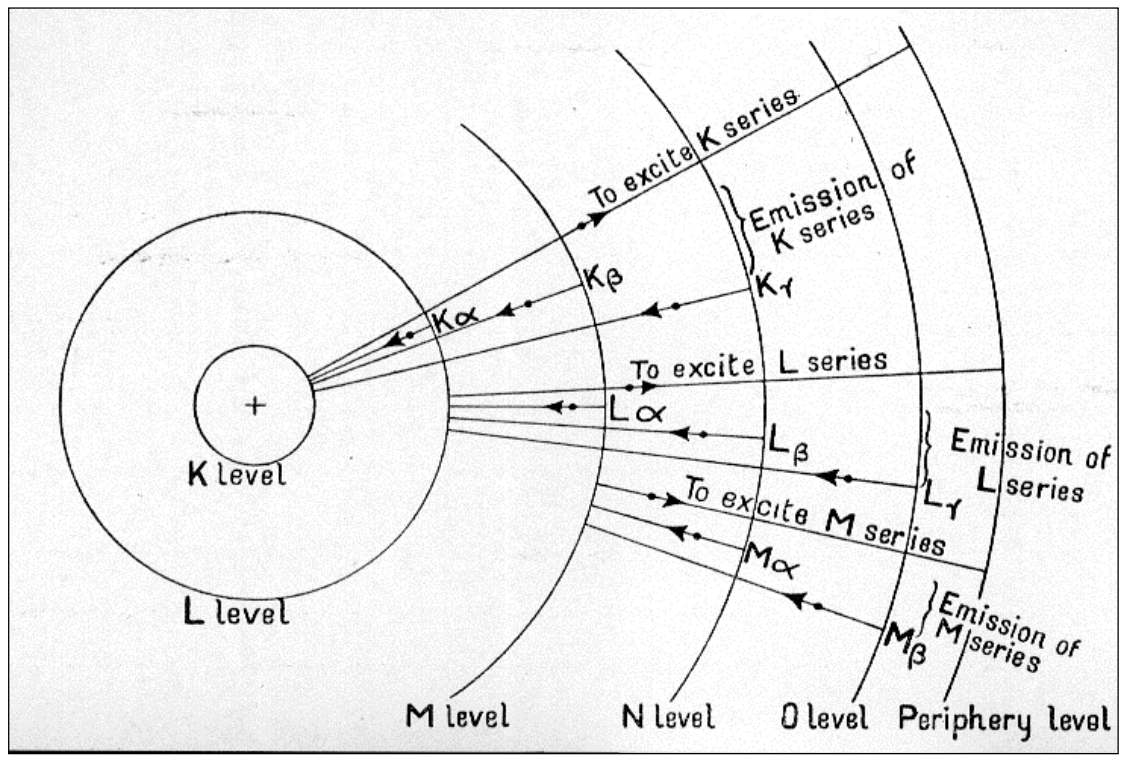

Figure 2. Scheme of energy levels of a heavy atom, with corresponding X-ray emissions. Source: E. N. da Andrade, *The Structure of the Atom*, G. Bell and Sons, London, 1923.

> This leads to the conclusion that the electrons, which are added further, should be put into concentric rings or shells, on each of which … only a certain number of electrons – namely, eight in our case – should be arranged. As soon as one ring or shell is completed, a new one has to be started for the next element; the number of electrons, which are most easily accessible, and lie at the outermost periphery, increases again from element to element and, therefore, in the formation of each new shell the chemical periodicity is repeated.

Kossel's table of the chemical elements gave, for the first time, the correct atomic numbers for all the known elements from hydrogen to uranium. Moreover, he provided population numbers for the shells of the lighter elements (up to $Z = 25$) that improved on those tentatively proposed by Bohr in 1913. For example, while Bohr had proposed (8, 2, 2) and (8, 8, 2, 2) for magnesium and calcium, respectively, Kossel argued that the two elements were filled with electrons according to (2, 8, 2) and (2, 8, 8, 2).

### 4. X-ray Atoms

Making use of a more advanced version of Kossel's reasoning, in 1917 Peter Debye at Göttingen University suggested a ring model based on the frequencies of the characteristic X-rays. Debye argued that the frequency due to an electron transition to the innermost K ring could be expressed as the energy difference between two rings, the energy depending on the number $p$ of electrons in the K ring. Ignoring outside influences, each of the K electrons experiences a central charge $(Z - s_p)e$, where $e$ is the elementary charge and $s_p$ is a screening effect caused by the other $(p - 1)$ electrons. The $K_\alpha$ transition will occur when one of the K electrons is removed to the L ring and then passes from this ring to the K ring. Debye showed that on these assumptions it followed from the Bohr-Kossel theory that

$$\frac{\nu(K_\alpha)}{R} = p(Z - s_p)^2 - (p-1)(Z - s_{p-1})^2 - \frac{(Z - p + 1)^2}{2^2},$$

where $R$ is the Rydberg constant. By fitting the $\nu(Z, p)$ function to the measured $K_\alpha$ frequencies for elements between $Z = 11$ (sodium) and $Z = 60$ (neodymium) he found good agreement for $p = 3$. Debye thus pictured the first electron ring as three symmetrically arranged electrons rotating around the nucleus. "From this ring one electron can be removed and be brought on a circular orbit associated with two quanta," he wrote. "The two remaining electrons then come closer to the nucleus and describe, at an angular distance of 180° from each other, a new circular orbit around the nucleus. The transition of of the three electrons from the second state to the first state creates the $K_\alpha$ line" [17].

The approach pioneered by Debye was refined by several other researchers, including Arnold Sommerfeld in Munich, Jan Kroo in Warsaw, and Vegard in Oslo. Kroo considered an atom with $(p - 1)$ electrons in the K ring and $q$ electrons in its L ring [18]. An electron would pass from K to L, leaving the two rings with $p$ and $(q - 1)$ electrons, respectively. From this picture he obtained an expression in which both $p$ and $q$ appeared:

$$\frac{\nu(K_\alpha)}{R} = p(Z - s_p)^2 - (p-1)(Z - s_{p-1})^2 + (q-1)\frac{(Z - p - s_{q-1})^2}{2^2}$$
$$- q\frac{(Z - p + 1 - s_q)^2}{2^2}$$

The $s$-quantities are screening factors due to the repulsions from the other electrons. Planck's constant does not appear explicitly in either Debye's or Kroo's expressions, but only implicitly in Rydberg's constant $R$. In fact, the only way in which quantum theory entered the derivation was through the quantization of the angular momentum. From a comparison of his formula with experimental data Kroo concluded that in the normal state $p = 3$ and $q = 9$. When an electron was lost by ionization, the atom would have $p = 2$, and after the emission of a $K_\alpha$ ray it would end in a state with $p = 3$ and $q = 8$.

Vegard had taken an interest in the Bohr atom at an early date [19], and in a series of works from 1917-1919, published in both German and English journals, he dealt extensively with atomic models derived from X-ray spectroscopic data. In November 1917 he concluded that his results agreed with experiments if elements with $Z > 9$ contained one K ring with quantum number $n = 1$ containing 3 electrons, two closely spaced L rings with $n = 2$ containing 7 and 8 electrons, respectively, and one M ring with $n = 3$ containing

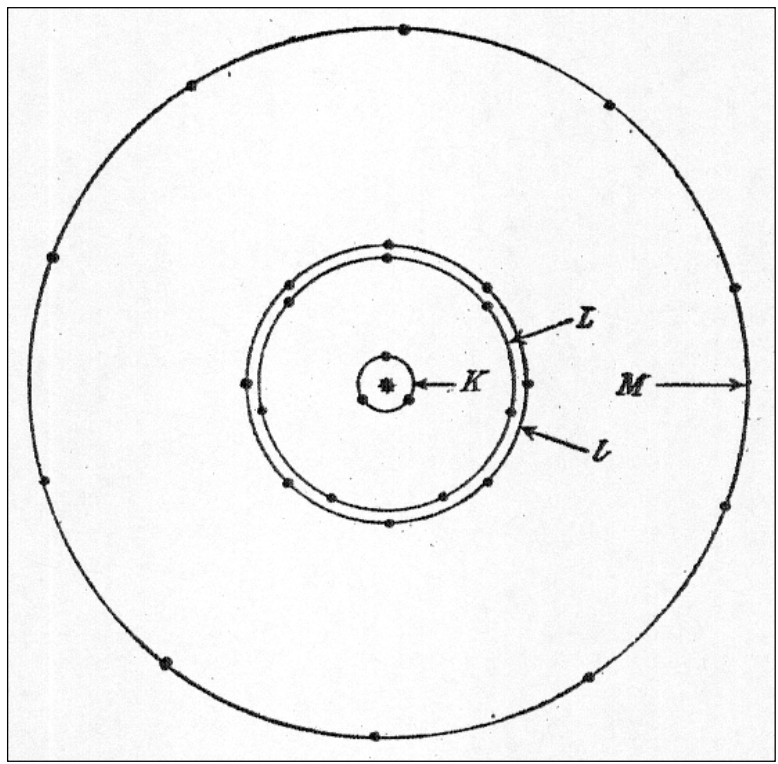

Figure 3. Vegard's model of the ring structure of the manganese atom [20].

9 or 10 electrons. Two years later he suggested that the best data indicated an M ring with twelve electrons, as shown in Figure 3 [20].

Vegard believed that there was an "*l* ring" just outside the L ring and that it had the same quantum number, $n$ = 2. For the light elements from lithium to fluorine he argued that they had an internal K ring of 2 electrons and an external L ring system increasing from 1 to 7 electrons: Li = (2, 1) to F = (2, 7). At neon the next electron would be added to the K ring rather than the L ring, meaning that Ne = (3, 7). He did not justify this configuration either theoretically or empirically, except that sodium was the first element for which K-radiation had been observed. For chemical reasons the structure (2, 8), as assumed by Kossel, might seem more reasonable, but Vegard did not comment on the discrepancy. In the period starting with sodium (3, 7, 1) the *l* ring would be built up, so that argon was assigned the electron structure (3, 7, 8). The next inert gas, krypton, was similarly characterized by an outermost ring containing 8 electrons, the structure being Ar = (3, 7, 8, 10, 8).

Whatever the population numbers, it is worth noticing that Vegard based his system on the assumption that the quantum numbers of the rings in the normal (unexcited) atoms increase by one unit as one moves outward from the nucleus. Whereas Bohr had assumed that the angular momentum of each electron in a many-electron atom was $h/2\pi$ (where $h$ is Planck's constant), according to Vegard it was given by $nh/2\pi$, where $n$ is the ring number. "I have succeeded," he wrote to Bohr, "to obtain a most striking agreement with experimental data on the basis of the hypothesis of increasing quantum numbers" [21]. Vegard's hypothesis implied that all elements belonging to the same period have the same value of the principal quantum number $n$. "If at all we shall be able to proceed further in the direction pointed out by Bohr," he said, "we can hardly avoid the assumption that systems of electrons exist in the normal atom with quant numbers greater than 1" [19]. That is, contrary to Bohr's original atom, which in its normal state was characterized by $n = 1$, Vegard's was a many-quantum atom.

By 1918 it was realized that there were two kinds of L orbits, either circular or elliptical. Both orbits had $n = 2$, but whereas the circular case was characterized by an azimuthal quantum number $k = 2$, the elliptical orbit had $k = 1$. (The azimuthal quantum number $l$ used in the later quantum mechanics is given by $l = k – 1$.) In order to place several electrons symetrically on the ellipse, Sommerfeld suggested in 1918 that each electron moved separately on its own ellipse, in such a way that at any moment each of the electrons would be at a corner of a regular polygon. What Sommerfeld referred to as an *Ellipsenverein* (union of ellipses) was adopted by Vegard in his theory of the elements, as shown in Figure 4. As he expressed it, "the elliptic axes are arranged radially and with equal angular intervals, and … at any moment the electrons will be evenly distributed on the circumference of a circle, the radius of which undergoes periodic changes as time passes" [22]. However, in his reconstruction of the periodic system he relied only on the principal quantum number $n$.

Debye, Sommerfeld, Kroo, and Vegard all agreed that, in the case of the heavier elements, the K ring contained three electrons, such that, for example, chlorine was assigned the structure (3, 7, 7) and phosphorus (3, 7, 5). Of course, these structures disagreed with the periodic system and other chemical knowledge. Nonetheless, for a few years they were widely accepted by the physicists, if not by the chemists. "Aren't there three electrons in the K ring?" a somewhat surprised Sommerfeld asked in 1919, when he realized that this might not be the case and that the ring atom might have to be abandoned [23].

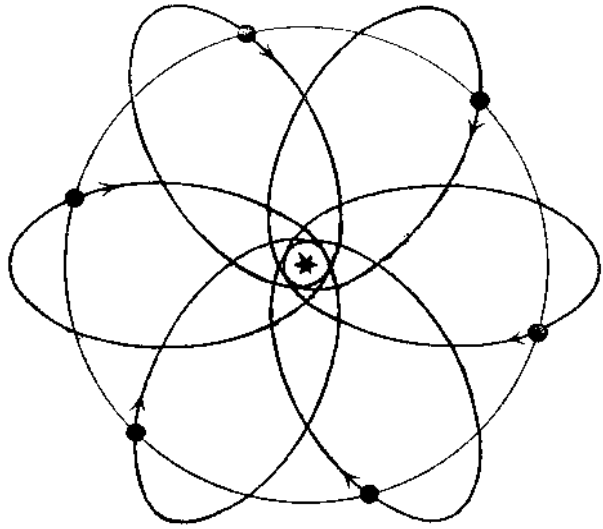

Figure 4. Vegard's illustration of electrons moving elliptically, yet in such a coordinated way that they are always placed on a circle. Source: L. Vegard, *Stoffets Opbygning og Atomenes Indre*, Olaf Norlis Forlag, Kristiania, 1924.

## 5. Vegard's Theory of the Periodic System

On the basis of X-ray data and his version of Bohr's ring atom Vegard attempted to provide all the atoms with quantum and population numbers, thus to account for the entire periodic system in terms of atomic theory. As mentioned, Bohr and Kossel had earlier made attempts in the same direction, but Vegard's project, as he presented it in two large papers in *Philosophical Magazine*, was more ambitious. His system was based on the general rule that the quantum number $n$ remained constant for all elements belonging to the same period, and that the value of this number also gave the order of the rings surrounding the nucleus. In his paper of 1918 he accounted for the heavier elements with atomic numbers in the interval $17 < Z < 55$ as follows [24]:

> From Ar we have both L-rings with 7 and 8 electrons formed… Now we come to the long periods from Ar to Kr. At first a ring of 10 electrons is formed, completed by the elements Fe, Co, and Ni with 8, 9, and 10 electrons in the external ring respectively; this should be the first M-ring with quant number 3. At Cu a new ring comes into existence, and we get a monovalent electropositive element. During the next long period from Kr to Xe the same process is repeated.

The rare earth elements were notoriously difficult to incorporate in a definite way in the periodic system [25], but in accordance with Bohr's suggestion of 1913 Vegard argued that they could be understood as elements in which a new ring with $n = 4$ was formed inside the outermost ring. He pictured Ba as (Xe, 1), meaning a xenon structure with one electron added in an external ring, and Ce as (Xe, 4). Passing to the next elements, "we assume the external ring to be kept, and that the new electrons are forming a new *internal* ring. … Thus the new electrons which are taken up in the series of the rare earths when we pass to higher atomic numbers are, so to speak, soaked into the atom, and the surface systems mainly determining the chemical properties are kept unaltered. How these new internal electrons are arranged we do not know." Vegard did not specify the number of rare earth elements, but from his periodic system (Figure 5) it appears that he included the still unknown element $Z = 72$ (hafnium) as a rare earth, thus assuming a series of 15 elements.

     As is well known, the question of the position of element 72 in the periodic system became a matter of controversy in 1922-1923 after Bohr had given theoretical reasons that it could not be a rare earth metal. According to Bohr the element must have an outer atomic structure similar to the one of zirconium, namely, two $n = 6$ electrons and ten $n = 5$ electrons. The rare earth group could only comprise 14 elements. In agreement with Bohr's prediction, hafnium was discovered by means of X-ray analysis in Copenhagen in December 1922 [26].

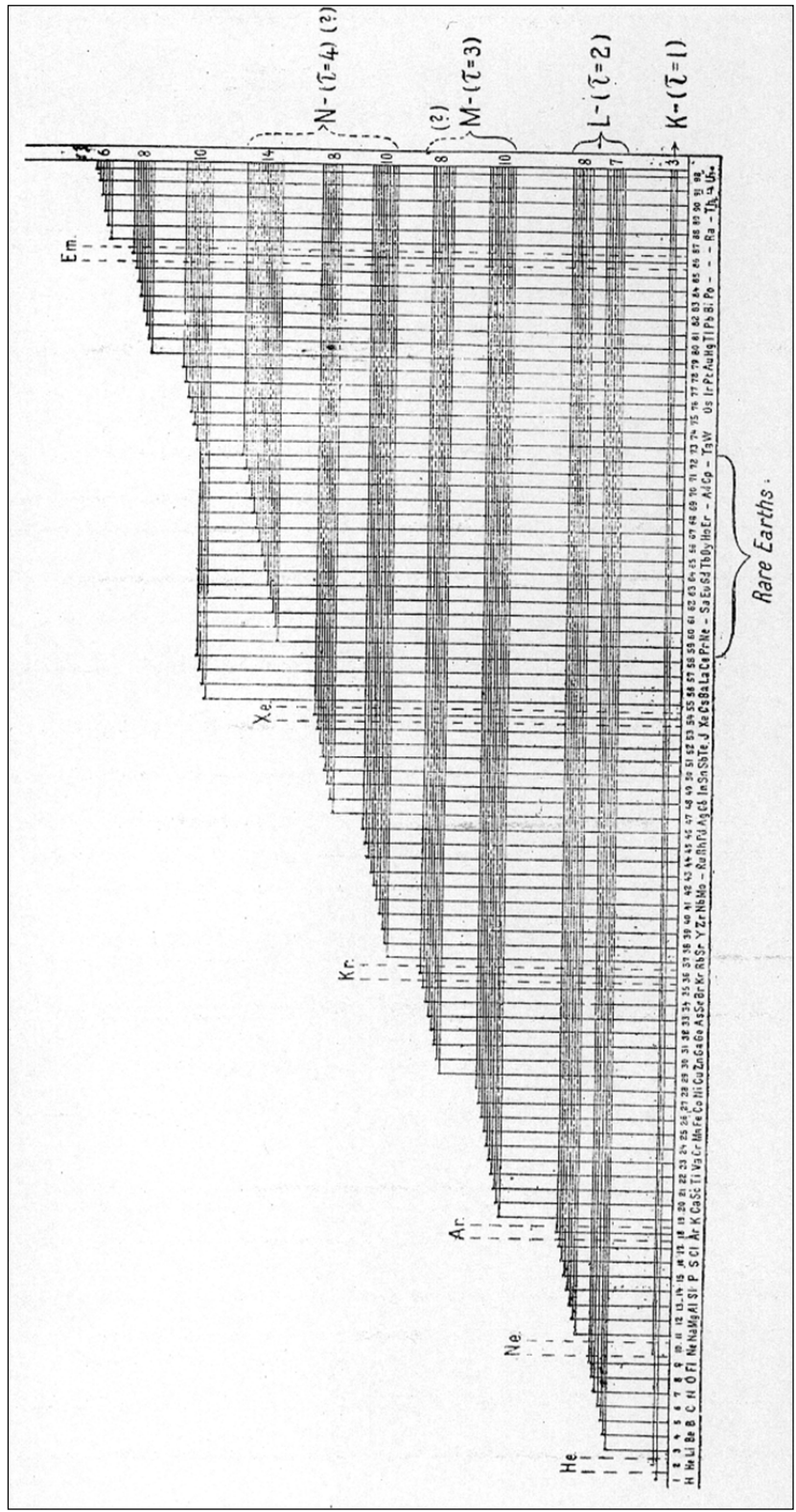

Figure 5. Vegard's graphical illustration of 1918 of the periodic system, with groups of electrons represented by horizontal lines. The electron arrangement of an element is obtained by drawing a vertical line from the place of the element on the horizontal axis.

In 1919 Vegard modified some of the results he had announced the year before, now assuming only a single L ring. He considered it certain that the K ring contained 3 electrons, and that the L ring comprised 7 electrons, whereas the assumption of 12 electrons in the M ring was seen as more uncertain. When it came to the higher atoms his population numbers were little more than educated guesswork. "We have more or less to grope in the darkness and feel our way forward," he admitted [27], and this he did by considerations of the same kind that had guided Bohr, Kossel and other atom-builders, that is, by taking into regard empirical facts concerning the chemical and physical properties of the elements.

One of those facts, used by Bohr and several earlier scientists, was Lothar Meyer's old curve showing the periodicity of the atomic volumes. Another and more recent empirical fact that Vegard made use of was the variation of atomic electrical conductivities with the atomic weight, such as shown in a curve published by the Swedish physicist Carl Benedicks [28]. According to Vegard, his theory of the periodic system was in striking agreement with Benedicks's curve.

Although Vegard expressed faith in his hypothetical explanation of the periodic system, naturally he was aware of its incompleteness and tentative character. Thus, he realized that he had not taken in to account interactions between the rings in his description of the atoms. At the end of his paper of 1919, he wrote: "We may also imagine a mutual connexion between the motions of the various ring systems. Now it is quite possible that these mutual relations may modify the properties of the atoms both as regards spectra, chemical, and physical properties" [29]. Indeed, as Bohr showed a few years later, the details of the periodic system could only be explained on the basis on the orbital atomic theory if the interaction between the orbits was taken into account.

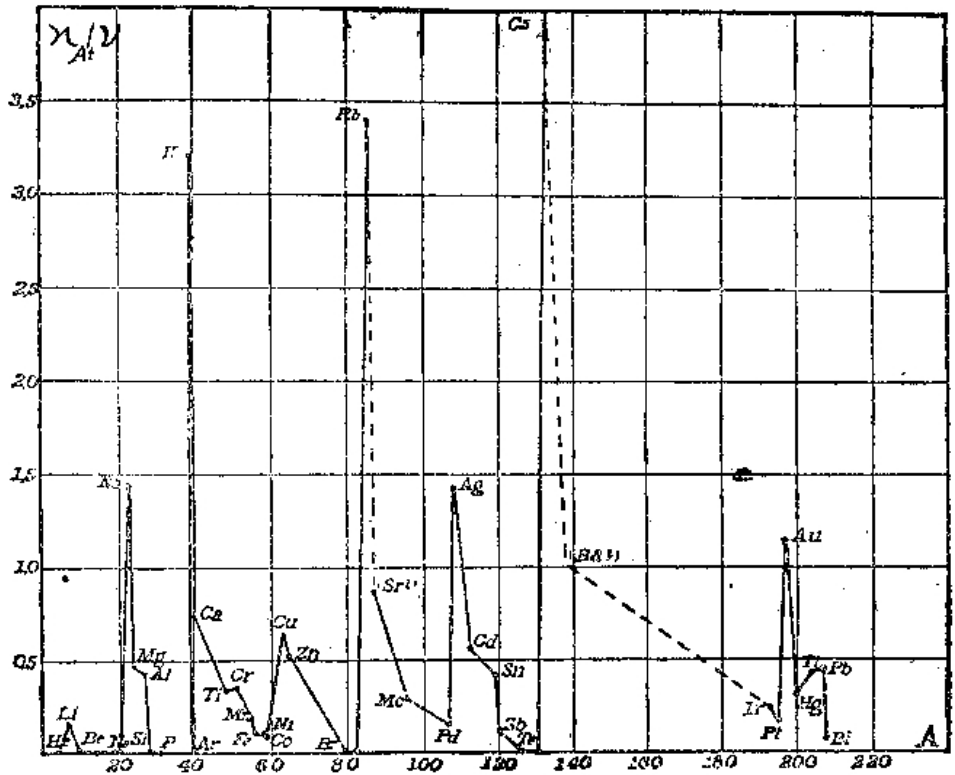

Figure 6. Benedick's curve of atomic electrical conductivities.

## 6. Reception and Later Development

Vegard's atomic theory and explanation of the periodic system was known in the chemical community [30], but without attracting much attention. Based as it was on lengthy calculations of atomic structure, it was not of a kind that appealed to the majority of chemists who favoured a more empirical approach. This approach was the one adopted by Irving Langmuir in his 1919 theory of atoms and molecules [31]:

> The problem of the structure of atoms has been attacked mainly by physicists who have given little consideration to the chemical properties, which must ultimately be explained by a theory of atomic structure. The vast store of knowledge of chemical properties and relationships such as it summarized in the periodic table, should serve as a better foundation for a theory of atomic structure than the relatively meager experimental data along purely physical lines.

Langmuir's theory, no less ambitious than Vegard's but building on an entirely different foundation, included a full periodic system with the number of electrons in the various shells. It did not refer to either X-ray calculations or Vegard's earlier theory. Among the few chemists who paid attention to Vegard's theory was Frederick Soddy, the chemistry Nobel laureate of 1921 for his contributions to radiochemistry. In a careful and sympathetic review of the theory, Soddy concluded as follows [32]:

> Without putting too much trust in the details of this theory of atomic structure, it presents us for the first time with a picture of the possible constitution of all the elements from one end of the periodic table to the other, which, however imperfect it may prove, is at least definite and capable of detailed quantitative examination and improvement as our knowledge of the high-frequency spectra of the elements grows.

Vegard's theory of the structure of atoms was short-lived and of limited influence on the process that led to an explanation of the periodic system. In his Nobel lecture of 1922, Bohr acknowledged two aspects of Vegard's work, namely, its explanation of the rare earth group and the idea of associating outer rings with a principal quantum number larger than one [18]. However, at the same time he noted that "Vegard's considerations do not offer points of departure for a further consideration of the evolution and stability of the groups, and consequently no basis for a detailed interpretation of the properties of the elements" [33].

A main problem of Vegard's theory was that it was based on the assumption of coplanar electron rings, which assumption soon turned out to be wrong. In a critical analysis of the Debye-Vegard approach, Fritz Reiche and Adolf Smekal demonstrated that Vegard's theory was unable to discriminate between, for example, population numbers (3, 7) and (2, 8) for the K and L rings; moreover, disturbances from one ring to another would spoil most of Vegard's results [34]. Reiche and Smekal consequently suggested that the planar ring atom might have to be abandoned and replaced by a structure in three dimensions. In a subsequent polemical publication Smekal reinforced his critique of Vegard's atom, which caused the Norwegian physicist to modify his model in a way which was, however, conspicuously ad hoc [35]. By 1921 Bohr, Sommerfeld and most other physicists abandoned the planar ring atom, and Vegard silently left atomic theory to work on what he considered his true vocation, the aurora borealis.

The X-ray approach cultivated by Vegard and other physicists turned out to be a blind alley. Instead, the main route that led to a full explanation of the periodic system in terms of atomic structure was a mixture of chemical considerations, as in the works of Charles Bury (1921) and John Main Smith (1923-1924), and methods largely based on quantum theory, as in Bohr's influential theory of 1921-1922 and the improved system that Edmund Stoner announced in 1924 [36]. Pauli's famous paper of 1925 [37], in which he introduced the exclusion principle as a theoretical foundation for explaining the periodic system, relied on the earlier works of Bohr and Stoner but only insignificantly on the chemical approach and not at all on Vegard's X-ray approach.